\documentclass[%
reprint,showkeys,
nofootinbib,  
amsmath,amssymb,
aps]{revtex4-2}

\usepackage{graphicx}
\usepackage{dcolumn}
\usepackage{bm}
\usepackage{siunitx}
\usepackage{xcolor}
\usepackage{balance}
\usepackage[english]{babel}

\begin{document}
	\title{How emulsified droplets induce the bursting of suspended films of liquid mixtures\\}

	\begin{abstract}
		 {Léa Delance,$^{1,2,3}$ Enric Santanach-Carreras,$^{3,4}$ Nicolas Passade-Boupat,$^{3,4}$ François Lequeux,$^{1,2}$ \\
		 	\indent Laurence Talini $^{5,*}$, Emilie~Verneuil $^{1,2,*}$\\
		
		{\it \noindent $^1$ Soft Matter Science and Engineering (SIMM), CNRS UMR 7615, ESPCI Paris, PSL University, Sorbonne Universit\'e, F-75005 Paris, France\\
			$^2$ Laboratoire Physico-Chimie des Interfaces Complexes, ESPCI Paris, 10 rue Vauquelin, F-75231 Paris, France\\
			$^3$ TotalEnergies S.A., P\^ole d’Etudes et de Recherches de Lacq BP47, 64170 Lacq, France\\
			$^4$ Laboratoire Physico-Chimie des Interfaces Complexes, Chemstartup, RD 817, 64170 Lacq, France\\
			$^5$ CNRS, Surface du Verre et Interfaces, Saint-Gobain, 93300 Aubervilliers, France\\
			$^{*}$ laurence.talini@cnrs.fr, emilie.verneuil@espci.fr\\}}
		
Emulsion droplets of silicone oil (PDMS) are widely used as antifoaming agents but, in the case of non-aqueous foams, the mechanisms responsible for the bursting of the films separating the bubbles remain unclear. We consider a ternary non-aqueous liquid mixture in which PDMS-rich microdroplets are formed by spontaneous emulsification. In order to quantitatively assess the effect of the emulsified microdroplets, we measure the lifetime of sub-micrometer-thick suspended films of these emulsions as well as the time variations of their thickness profiles. We observe that a droplet entering the film reduces its lifetime by inducing a local and fast thinning. In most cases, we ascribe it to the spreading of the drop at one of the film interface with air, which drags the underlying liquid and causes eventually the film to burst rapidly. We explain why, despite slower spreading, more viscous droplets cause films to burst more efficiently. More rarely, microdroplets may bridge the two interfaces of the film, resulting in an even more efficient bursting of the film that we explain.	
	\end{abstract}
	
	\maketitle
	

\section{Introduction}

\begin{figure*}[t!]
	\centering
	\includegraphics[width=0.9\linewidth]{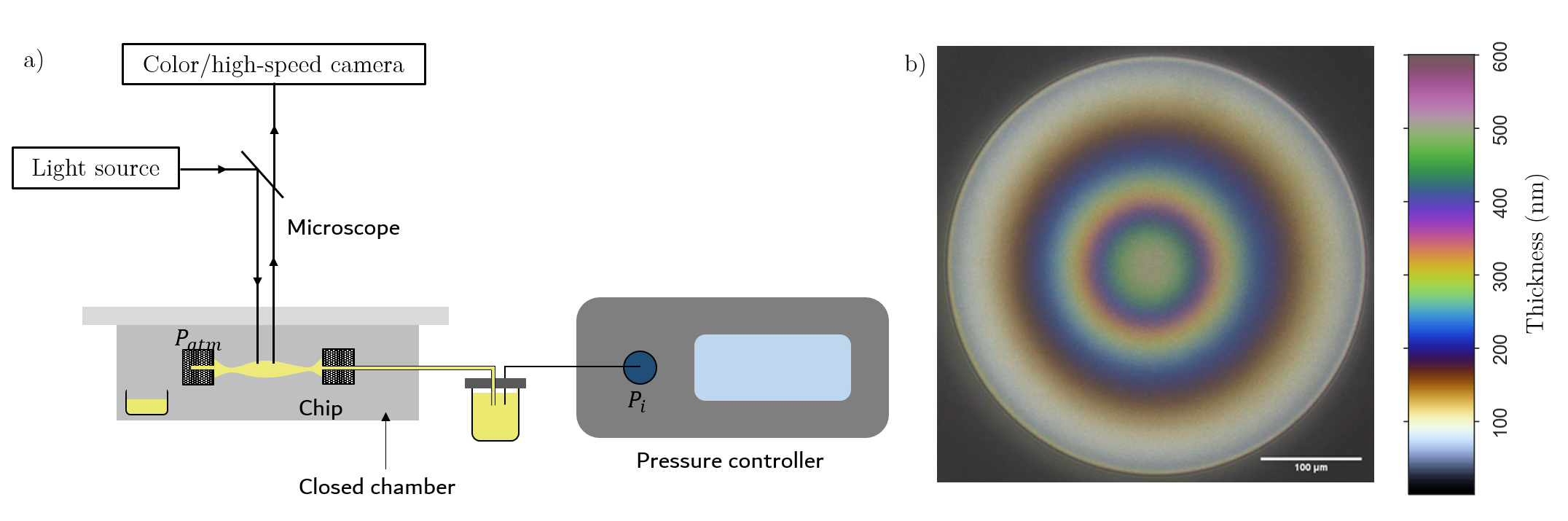}
	\caption{a) Schematic of the set-up used for suspended thin film experiment. The films are formed using a "bike-wheel" microfluidic chip connected to a pressure controller. They are illuminated with reflected light which allows for the measurement of the film's thickness over time and space by interferometry. b) Typical image obtained along with a color-to-thickness scale bar. The film is thicker at the center than on the edge, as a result of a stabilising mechanism arising from the surface elasticity of the oil mixture when confined in thin films \cite{tran_mechanisms_2022}. 
	}
	\label{fig:schema_manip}
\end{figure*}

Controlling the foamability of liquids is of great importance in many industrial processes ranging from gas-liquid separation in oil and gas production to paper or drugs production and wastewater management \cite{shaban_study_1995, patterson_influence_1993, yiengpruksawan_mucolytic-antifoam_1991}. Over the last decades, a common strategy to decrease the production of foam has been to disperse Polydimethylsiloxane (PDMS) as microdroplets in the liquids of interest \cite{kelland_production_2014}. Different mechanisms at the scale of the thin liquid films separating the bubbles have been proposed to explain the antifoaming effect of PDMS droplets \cite{pugh_foaming_1996,bergeron_polydimethylsiloxane_1997, denkov_mechanisms_1999,garrett_defoaming_2015}: the droplets {may} either spread at the surfaces of the films or bridge them. Then, either by spreading or stretching, {the films are made to thin down and eventually rupture}. Yet, there is a lack of experimental validation of these mechanisms. Denkov {\it et al.} \cite{denkov_mechanisms_1999} observed a bridging-stretching mechanism but, to the best of our knowledge, no  {direct observation} of a spreading mechanism has been found in antifoaming additives. A spreading mechanism was identified in quite different systems: oil droplets, with diameter of order 20~$\mu$m, were found to induce the bursting of expanding aqueous liquid sheets with thicknesses in the 10 to 100 $\mu$m range \cite{vernay_bursting_2015}. Moreover, these mechanisms have mostly been studied in the case of foaming aqueous {solutions}, in which the surfactants used to stabilise the thin films can interact in a complex way with the droplets, for example preventing their entry or their spreading at the air-liquid interface\cite{bergeron_polydimethylsiloxane_1997, denkov_role_2002, denkov_mechanisms_2004}. {As a result, the destabilising mechanisms due to microdroplets are expected to strongly depend on the continuous phase they are dispersed in,  {as stated by Denkov {\it et al.} \cite{denkov_mechanisms_1999} } }. \\
Hence, apart from water-based soapy liquids, foams may also form in surfactant-free systems, although with lifetimes at most of tens of seconds for water-like viscosity. It is expected \cite{garrett_defoaming_2015} that the absence of surfactants makes the entry of a drop at the air interface easier, thereby potentially increasing the anti-foaming efficiency of added microdroplets.  {Along this line, Chandran Suja \textit{et al.}\cite{chandran_suja_foam_2020} studied the antifoaming effect of droplets dispersed in lubricants, which are oil mixtures. They showed that the efficiency of the antifoaming agent decreases after the lubricants have been filtered several times
	and demonstrated that smaller sizes of dispersed droplets lead to longer bubble lifetimes. This size effect was also evidenced in water-based foams \cite{bergeron_polydimethylsiloxane_1997}.} However, further work is needed to provide a quantitative understanding of antifoaming effects in liquid mixtures. 
In this paper, we focus on continuous phases composed of organic liquid mixtures. In those surfactant-free liquids, the lifetimes of foams or bubbles are increased compared to that of pure liquids because of small concentration differences of their components existing between their bulks and their surfaces \cite{tran_understanding_2020, tran_mechanisms_2022}. 
Indeed, the one component with the lowest surface tension tends to adsorb more at the air interface and its interfacial concentration is slightly larger than the bulk one. When the film is very thin, its further thinning down at constant volume requires a change in both bulk and surface concentrations, resulting in a thickness-dependent interfacial tension. A depleted reservoir effect opposes further thinning which translates into a surface elasticity.

In the present study, we describe the mechanisms by which microdroplets dispersed in those mixtures accelerate the bursting of suspended thin films. To this end, we use well controlled surfactant-free continuous phases and droplets formed from ternary mixtures of liquids at compositions for which spontaneous emulsification occurs \cite{delance_uptake_2022}. We conduct statistically-rich measurements of the lifetime of thin films coupled with a fine characterisation of their thickness resolved over space and time using interferometry. \\

\section{Experimental section}

\subsection{Ternary liquid mixture}\label{sec:ternary_liquid_mixture}
We study films composed of a ternary system of Polydimethylsiloxane (PDMS, trimethylsiloxy terminated, from ABCR), cyclopentanol (purity$>$99\%, from Sigma-Aldrich) and decane (purity$>$99\%, from GPR Rectapur). Three different PDMS were used: they are characterised by their viscosity which is 0.5~Pa.s, 6~Pa.s and 30~Pa.s. The mixture consists in 70\% of cyclopentanol, between 400 and 1500 ppm of PDMS and the rest (about 30\%) of decane.  All fractions are indicated in mass. The viscosity of the decane/cyclopentanol mixture (30:70 w:w) is 3.5~mPa.s. To make the mixture, PDMS is first diluted in decane, then the mixture is added in cyclopentanol.
In a previous work \cite{delance_uptake_2022}, for these compositions, we have demonstrated that  the mixture separates into two phases through spontaneous emulsification. We obtain PDMS-rich microdroplets (60\% PDMS, 25\% decane, 15\% cyclopentanol) dispersed in a continuous phase composed of 70\% of cyclopentanol, 30\% of decane with traces of PDMS. By separating the two-phases, we have checked that varying the viscosity of the pure PDMS has little effect on the composition of the dispersed phase (see the Supplementary Information).  {The interfacial tension in air of the continuous and dispersed phases were also measured by the rising bubble method to be $\gamma_{c}=26.9$~mN.m$^{-1}$ and $\gamma_d=21.7$~mN.m$^{-1}$ } respectively. The newtonian shear viscosity $\eta_d$ of the  {dispersed} phase was systematically measured (see Supplementary Information) and is equal to 0.3, 6 and 27~Pa.s depending on the initial PDMS grade,  {always larger than that of the continuous phase $\eta_c=3.5$~mPa.s$^{-1}$}.\\
Over time,  {right after mixing, spontaneous emulsification leads to an increase in the radius $R_d$ of the emulsified microdroplets due to diffusion-limited coalescence (see SI)} and obeys the following power law:
\begin{equation}
R_d\simeq R_0\left(\frac{t}{\tau_c}\right) ^{1/3}.
\label{eq:coalescence}
\end{equation}
with
\begin{equation}
\tau_c = \frac{2\pi \eta_c R_0^3 }{k_BT\phi},
\label{eq:tau}
\end{equation}
Note that the dependence in initial radius $R_0$ cancels out in Eq.~\ref{eq:coalescence}. Here, $\eta_c$ is the viscosity of the continuous liquid phase (see SI), $k_BT$ the thermal energy, and $\phi$ the droplet volume fraction. In our conditions, $R_0$ is typically 0.1~$\mu$m and $\tau_c\approx 2$~s. Here, by controlling the time $t$ between the mixing of the three liquids and the suspended film experiments, we work with well-known microdroplet radii $R_d$. The latter is adjusted by varying the volume fraction $\phi$: for $\phi=400, 800, 1500$~ppm, we obtain three values of $R_d$=0.7, 1.0 and 1.2$\pm0.3~\mu$m respectively. Note that the number density of microdroplets does not change with $\phi$ at a given time.

Moreover, we have demonstrated that these droplets enter without delay (no energy barrier except the hydrodynamic one) at an air-continuous phase interface and spread at this interface, forming a homogeneous thin PDMS-rich layer. 

In order to isolate the effect of the droplets on the stability of the films, we also study the mixture without droplets, i.e. only the continuous phase. To extract it, we wait for the droplets to sediment and we remove the upper phase, which we filter using a \SI{0.2}{\micro\meter} pore size PTFE filter to suppress any remaining droplet. This procedure ensures the continuous phase contains the same traces amount of solubilized PDMS chains. In the following, this solution will be referred to as the continuous phase as opposed to the emulsion.

\subsection{Suspended thin film experiment}

Using the mixtures described above, we then carry out thin film experiment with a set-up schematized in Fig.\ref{fig:schema_manip}. The films are made in a microfluidic chip made of fused silica with a bike-wheel design inspired from Cascão Pereira \textit{et al.}\cite{cascao_pereira_bike-wheel_2001}. The hole in which the film is suspended has a diameter of 1~mm. The liquid is brought to or sucked up from the hole through 36 cylindrical channels with a diameter of \SI{50}{\micro\meter}. The chip is put in a close chamber \cite{beltramo_millimeter-area_2016} with inner atmosphere at vapor-saturation with the liquid mixture in order to avoid evaporation. A pressure controller (OB1 MK3+, Elveflow, 0-200 mbar) sets an overpressure $P_i$ in the liquid reservoir. The level of the liquid in the reservoir is $\Delta H$=-2~cm below the height of the chip: this produces a hydrostatic pressure difference which allows to suck the liquid from the chip when $P_i=0$. The film is observed with a microscope (Zeiss) with episcopic illumination (white metal halide light source (HXP 120)), a x5 {objective (NA=0.16)} and a colour camera (exo252CU3, SVS-Vistek) {at 70~fps}. The hue is converted into thickness by interferometry after a careful white balance. A typical image is shown in Fig.\ref{fig:schema_manip}b along with the conversion scale between the hue and the film's thickness. We also observe the final film bursting using a high-speed camera at 76500 fps (Photron).

The experimental procedure was adapted from the literature \cite{chatzigiannakis_breakup_2020, chatzigiannakis_studying_2022} as follows. The chip is first completely filled with liquid and we ensure that there is no air bubbles altering the applied pressure. {The liquid is further injected in the hole until a swollen suspended drop is obtained. It is then sucked out by applying a lower pressure $P_i$ in the reservoir: a meniscus forms, then a thin liquid film appears at the center and thins.}
{
	Before the formation of the thin film and considering the capillary and hydrostatic contributions, the pressure in the liquid $P_l$ is:
	\begin{equation}
	P_l = P_{atm}+P_i- \frac{2\gamma_c}{r} -\rho g\Delta H,
	\label{eq:pressure}
	\end{equation}
	with $r$ the radius of curvature of the meniscus, taken as the cell hole diameter, $\rho$ the liquid density, and $g$ the gravitational acceleration.
	Prior to any series of measurement, we determine experimentally the pressure $P_{i0}$ needed to obtain a stable film, i.e. the pressure for which $P_l=P_{atm}$. Using Eq.\ref{eq:pressure}, we find that it corresponds to: $P_{i0}=\frac{2\gamma_c}{r}+\rho_lg\Delta H$. The pressure difference $\Delta P$ is applied compared to this reference pressure, meaning that $P_i = P_{i0}+\Delta P$.}

{In all experiments, a negative pressure $\Delta P$=-40~Pa is applied \cite{chatzigiannakis_studying_2022}.}  {The initially thick meniscus thins down until the first interference fringes appear, which we define as time $t=0$ \cite{chatzigiannakis_breakup_2020}. Images are recorded until the film bursts. The film's lifetime is defined as the time difference between the moment of bursting and time $t=0$.} The profile of the film, defined as its half-thickness as a function of the distance to its center, is measured over time by interferometry with a resolution of $\pm 4~$nm {assuming a planar geometry and a 0° incidence angle. We have verified that the latter assumption holds, considering the numerical aperture of the used objective (0.16) and the very small slope of the interfaces with air ($10^{-3}$). \cite{limozin_quantitative_2009}}  In order to renew the liquid in the chip, a few drops are expelled and removed from the chip after 5 consecutive experiments.

\section{Results}

\subsection{Films composed of the continuous phase}
\label{continuous_phase}

\begin{figure}[h]
	\centering
	\includegraphics[width=1\linewidth]{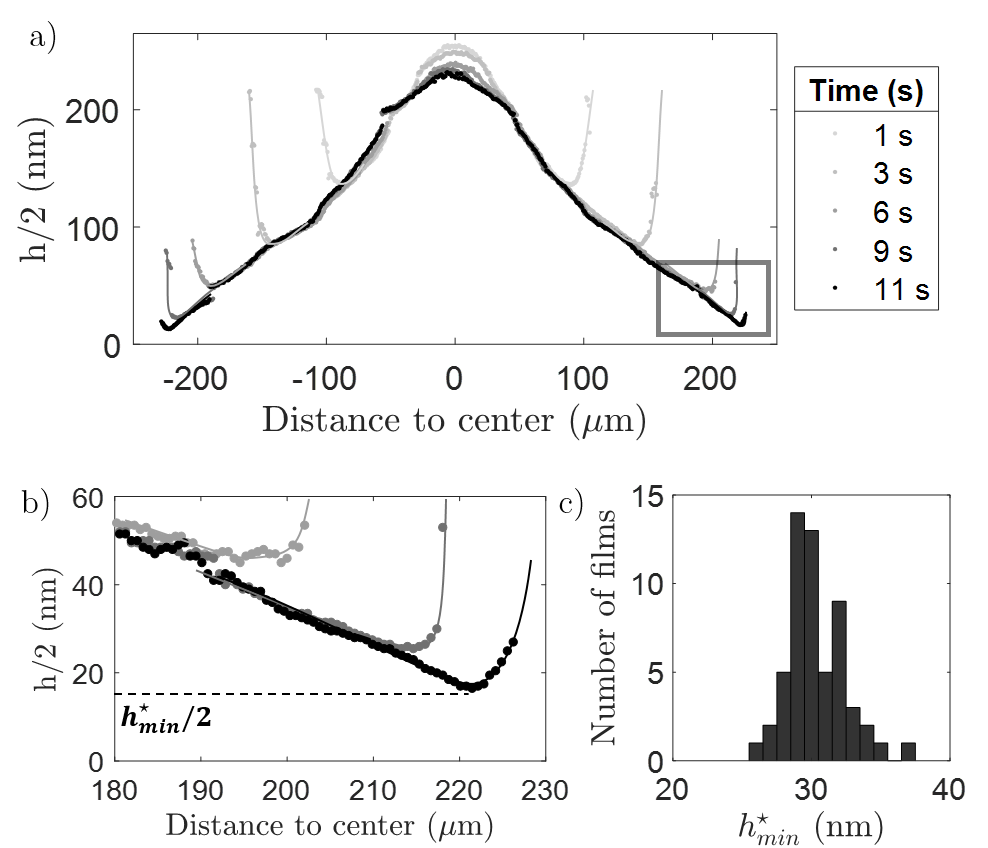}
	\caption{(a) Profiles over time of a typical film composed of the continuous phase of the emulsion. Legend: elapsed time since the formation of the film (in seconds). A typical image from which these profiles are obtained is shown in Fig.\ref{fig:schema_manip}. The film being axisymmetric, the profiles are measured on a single line crossing the center. The profile at 11~s corresponds to the last image recorded before the bursting of the film.
		b) Enlarged view of the dimple's neck. At the time of bursting, the minimal thickness is denoted $h*_{min}$.
		c) Distribution of the films' minimal thickness before bursting over 60 films: $h*_{min}$=30 nm$\pm$2~nm.}
	\label{fig:phase_continue}
\end{figure}

\begin{figure*}[t]
	\centering
	\includegraphics[width=0.9\linewidth]{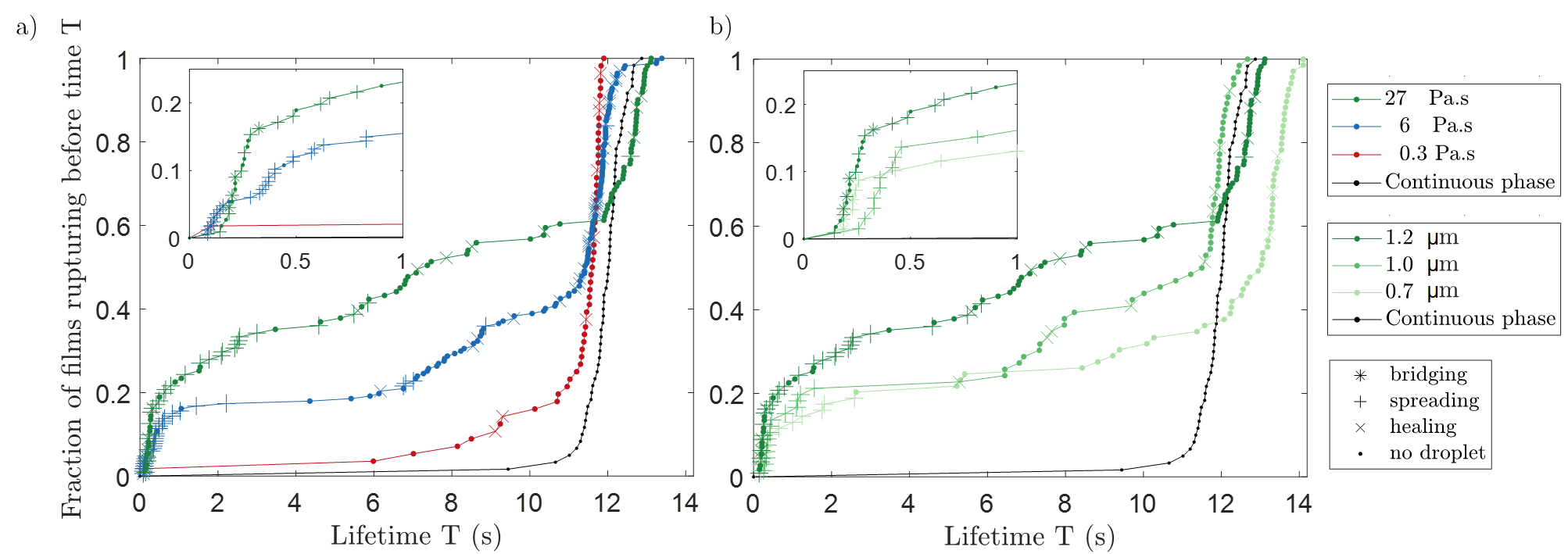}
	\caption{Cumulative distribution of the film's lifetime depending on (a) the viscosity $\eta_d$ and (b) the radius $R_d$ of the droplets. Black curve : films only composed of the continuous phase of the emulsion without droplets. Insets : close-up view on lifetimes shorter than 1~s. (a) PDMS concentration= 1500 ppm, $R_d=1.2\pm 0.3~\mu$m. 
		Except for the least viscous PDMS, lifetime distributions are bimodal with short lifetimes of about 0.3~s and longer lifetimes equal to the lifetime of films composed of the continuous phase. The number of short lifetimes increases with the viscosity of the droplets. (b) Droplet viscosity $\eta_d=27~$Pa.s. The ratio of short lifetimes increases with the size $R_d$ of the microdroplets.}
	\label{fig:distributions}
\end{figure*}

Firstly, we focus on films only composed of the continuous phase. The profile over time of a typical film is shown in Fig.\ref{fig:phase_continue}. We observe that the film is thicker at the center than on the edge. Over time, the thin film expands and gets thinner on the edge (Fig.~\ref{fig:phase_continue}b). It bursts as soon as it reaches a critical thickness $h_{min}^\star$ of $30\pm2$~nm with an excellent repeatability, the uncertainty corresponding to the standard deviation computed over 60 films (Fig.~\ref{fig:phase_continue}c). This dimple shape is usually observed in thin liquid films stabilised by surfactants. In our experiments, it results from the surface elasticity of the oil mixture, whose origin has been described earlier \cite{tran_understanding_2020, tran_mechanisms_2022} : the small concentration differences existing between the bulk and the surface of a binary mixture results in Gibbs' elasticity which sustains surface tension gradients between the thin film and the meniscus. In the present work, we do not further detail this mechanism.

Statistical measurements of the film's lifetime are reported as a black line in Fig.\ref{fig:distributions} and confirm the existence of a stabilising effect. We find a unimodal distribution with a mean lifetime of $12\pm0.5$~s. This indeed is much larger than the lifetime of films composed of pure liquids of same viscosity that is of order $\eta_c d/4\gamma_c\approx 10^{-5}~$s \cite{lhuissier_bursting_2012, debregeas_life_1998}. {This estimate arises from a balance between capillary pressure $4\gamma_c/d$ and viscous dissipation as the film drains by stretching due to free slipping conditions at the film interfaces.}

\subsection{Emulsion films: three different behaviours}

We now focus on the behaviour of emulsion films and the effect of microdroplets on their lifetimes. As detailed in Section~\ref{sec:ternary_liquid_mixture}, experiments were performed with different emulsions characterized by their microdroplet radius $R_d$, i.e. 0.7, 1.0 and 1.2$\pm0.3~\mu$m, as well as the viscosity of the PDMS-rich dispersed phase $\eta_d$.\\

To do so, we first review the behavior of films of the same emulsion, which contains 1500 ppm of PDMS of intermediate viscosity, so that $R_d=1.2~\mu$m and $\eta_d=6$~Pa.s. Three different behaviours are observed in the films of those emulsions, which are described below.

\begin{figure*}[t!]
	\centering
	\includegraphics[width=1\linewidth]{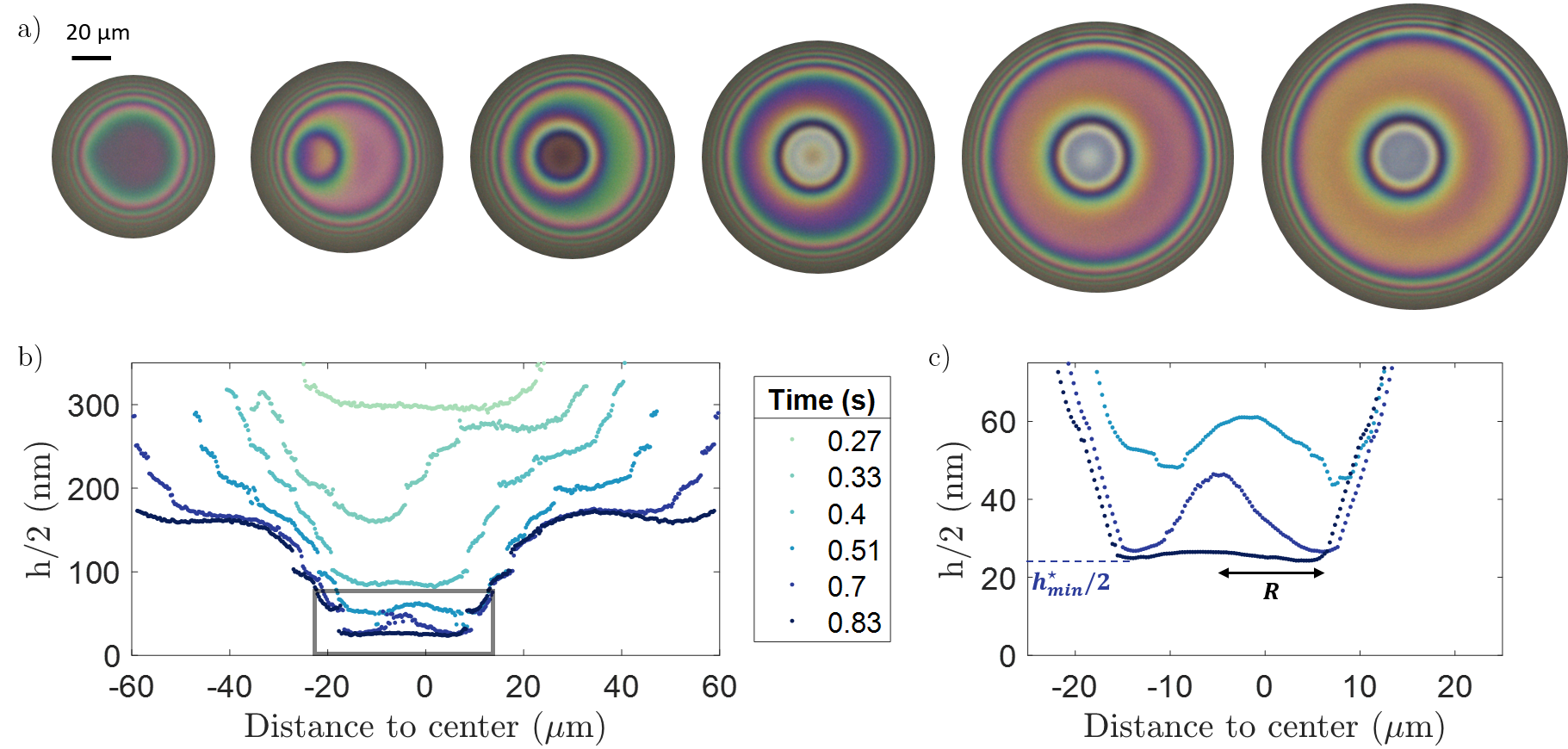}
	\caption{Film in which a droplet causes the bursting: spreading. (a) Time-lapse images: as the neck of the dimple moves outward, a depression forms at the center of the film and thins down. See Fig.~\ref{fig:schema_manip} for hue scale. (b) Corresponding profiles measured on a line crossing the center of the depression and the center of the film. (c) Enlarged view of the profiles at the location of the depression at late times evidencing a sub-dimple. Right before bursting, depression thickness denoted $h^*_{min}$ and radius $R$. Lifetime = 0.83~s. Droplet radius $R_d$=\SI{1.2}{~\micro\meter} and viscosity $\eta_d$=6~Pa.s.}
	\label{fig:goutte_spreading}
\end{figure*}

\begin{figure}[h!]
	\includegraphics[width=1\linewidth]{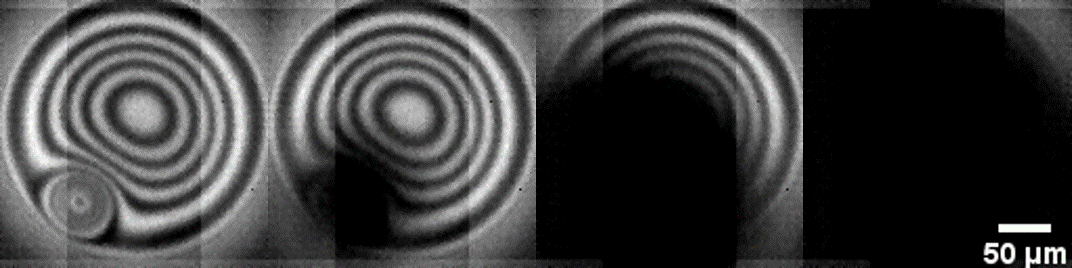}
	\caption{Time-lapse images of a film having a depression on its edge : bursting nucleates from the location of the depression. Successive images are separated by 13~$\mu$s.  }
	\label{fig:bursting:rapide}
\end{figure}

\begin{figure}[h!]
	\centering
	\includegraphics[width=1\linewidth]{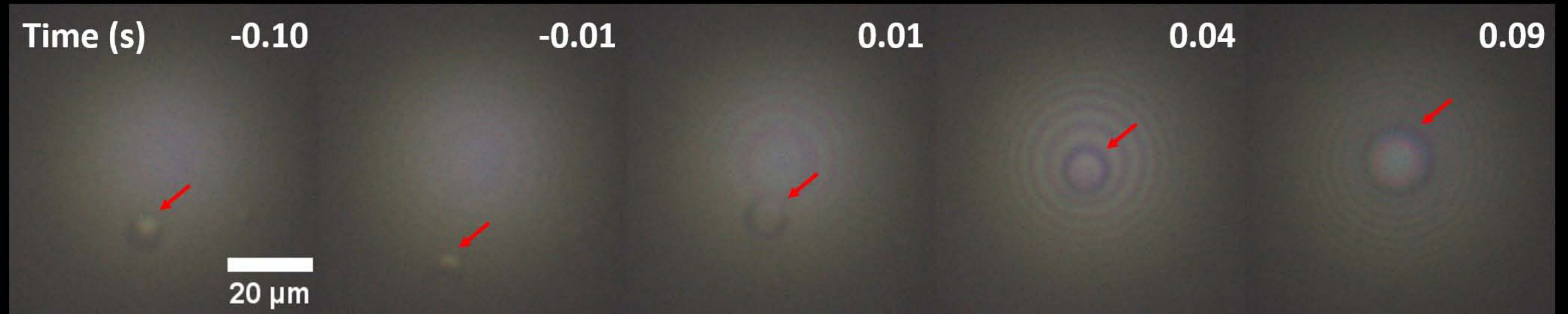}
	\caption{Film in which a droplet causes the bursting: bridging. A microdroplet is marked by a red arrow. Times t=-0.1 and -0.001~s: slowly expelled out of the film. At t=0.01~s, sucked toward the center. At t=0.04 and 0.09~s, its diameter finally increases rapidly. The bursting of the film occurs at the microdroplet location. Lifetime = 0.09~s. Droplet radius $R_d$=\SI{1.2}{~\micro\meter} and viscosity $\eta_d$=6~Pa.s.}
	\label{fig:goutte_bridging}
\end{figure}

A first behaviour is characterised by the formation of a depression in the film and the subsequent fast bursting of the film. An example is shown in Fig.\ref{fig:goutte_spreading}a-c. On the images and profiles, we observe a depression forming on the side of the film and moving toward its centre. The enlarged view at the location of the depression Fig.~\ref{fig:goutte_spreading}-c shows a thicker zone inside the depression early on, which flattens over time. We will refer to this secondary structure as a sub-dimple, and we will discuss its formation in the next section. The film bursts when its minimal thickness $h_{min}^\star$ reaches $50\pm 4~$nm. High-speed camera images in Fig.~\ref{fig:bursting:rapide} reveal that bursting occurs at the location of this depression.
Furthermore, a careful examination of the images reveals that a depression occurs when a droplet was visible before the first interference fringes appear. Also, the depression forms on the side of the dimple where the microdroplet was first seen. 
Hence, in most cases, droplets come from the meniscus and enter the dimple as the latter expands radially. In this example (Fig.~\ref{fig:goutte_spreading}~), the film has a relatively short lifetime of 0.83~s. A microdroplet can also be seen in the other example shown in Fig.\ref{fig:goutte_bridging}: a droplet (indicated by the red arrow) first moves away toward the meniscus and is then trapped in the center of the film. Afterwords, the apparent radius of the droplet increases, thus revealing its shape departs from a sphere. The film finally bursts quickly: it lasts only 0.09~s compared with the example in Fig.~\ref{fig:goutte_spreading} which has a lifetime of 0.83~s.

\begin{figure*}[t!]
	\centering
	\includegraphics[width=1\linewidth]{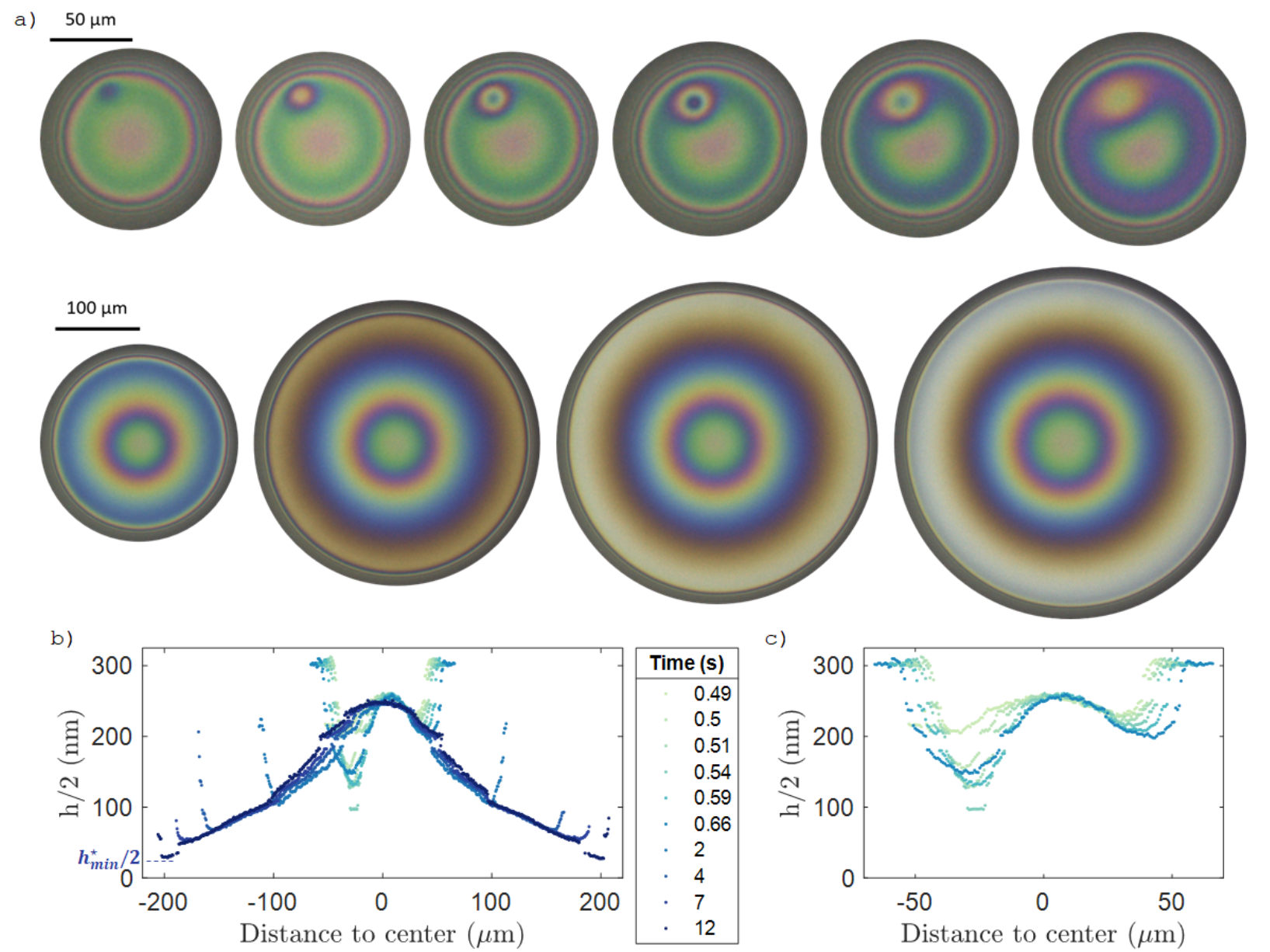}
	\caption{Images (a) and corresponding profiles (b) of a film in which a depression heals. See Fig.~\ref{fig:schema_manip} for hue scale. Profiles are measured along a diameter of the film also crossing the center of the depression. (c) Enlarged view of the profiles at the times when the depression forms and heals. Droplet radius $R_d$=\SI{1.2}{~\micro\meter} and viscosity $\eta_d$=6~Pa.s}
	\label{fig:disparition_goutte}
\end{figure*}

We now turn to the second type of behaviour encountered with emulsion films, which is illustrated in Fig.\ref{fig:disparition_goutte}. As shown, a depression forms between 0 and 0.54~s but, in contrast to the former behaviour, it heals between time 0.54 and 2~s. The film subsequently thins down similarly as dimpled films formed using the continuous phase. In the given example, the thickness reached at the center of the depression is 200~nm, which is much larger than the critical thickness $h_{min}^\star$  for film bursting. Eventually, $h_{min}^\star$ is reached after 12~s at the neck of the dimple.

The third and last observed behaviour corresponds to films in which no depression is formed. The thinning is similar to the one of the films composed of the continuous phase (Fig.\ref{fig:phase_continue}), except that the critical thickness $h_{min}^\star$ is larger ($65\pm2~$nm), again with a good reproducibility. \\

Depending on the experimental conditions (droplets radius and viscosity), relative occurences of different behaviours vary  {(see histograms in SI)}. The most likely (between 60\% and 80\% of the films) is the one for which there is no depression. Depressions leading to bursting are observed for 2\% to 30\% of the films while the cases in which a depression heals represent between 5\% and 20\% of the films. 

In this section, we have seen that PDMS-rich microdroplets dispersed in a liquid mixture can induce the early bursting of suspended thin films via the formation of depressions. However, these depressions do not systematically induce film bursting since they may spontaneously heal. In the next section, we study the effect of the droplet radius and viscosity on the distributions of the film lifetimes.

\subsection{Effect of droplets viscosity and radius}

In Fig.\ref{fig:distributions}, we present statistical measurements of film lifetimes for droplets of different viscosity and radius. 
In contrast to the unimodal distributions observed with films of the continuous phase, the films formed from the emulsions exhibit bimodal distributions of lifetimes with short times (less than 2 seconds) corresponding to the films for which a depression leads to bursting and longer times corresponding to the two other cases (healing of the depression or no depression). We emphasize that the longer lifetimes (12 seconds) are equal to the mean film's lifetime found for the continuous phase. We also observe intermediate lifetime values. They correspond either to films in which a depression enters late or to films that burst early without any depression formed.

We find that the ratio of short lifetimes increases as the viscosity of the droplets increases. Interestingly, we observe almost no short lifetimes for the droplets with the smallest viscosity. Moreover, we show that increasing the radius of the droplets accelerates the bursting of films.

In the next section, we aim at explaining the reasons for the formation of these depressions, whether they induce the bursting of the film or heal, and analysing the effect of droplets viscosity. 

\section{Discussion}

\subsection{Mechanisms of film bursting in oil foams}
\label{discussion_1}
In the following, we focus on the films for which a depression leads to its bursting and we explain why these depressions form. 

\begin{figure}[h!]
	\centering       \includegraphics[width=0.9\linewidth]{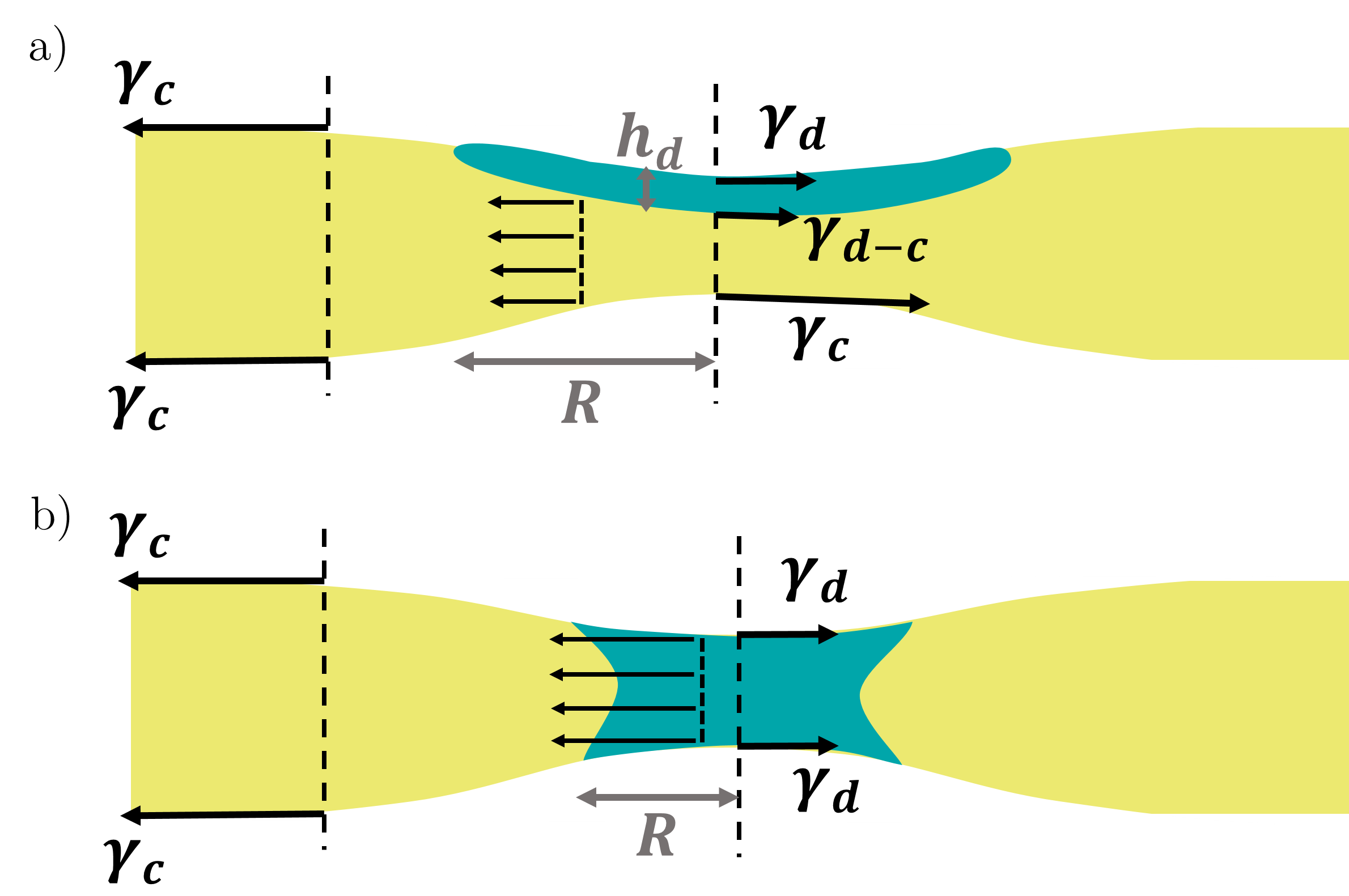}
	\caption{Schemes summarising the mechanisms of destabilisation of thin aqueous films of thickness $h$ by microdroplets. A droplet either (a) spreads at the surface of the film, or (b) bridges the film and stretches. Droplet assimilated to a disk with final radius $R$ and thickness $h_d$: in case (b), $h=h_d$. The local imbalance $S$ of surface tensions drives the thinning of the film and finally its bursting (a) $S=\gamma_c-(\gamma_d+\gamma_{d-c})$; (b) $S_{b}=2(\gamma_c-\gamma_d)$.}   \label{fig:schema_spreading_bridging}
\end{figure}

The destabilisation of thin films by droplets or particles has been studied in the literature mostly in the case of aqueous foams stabilised by surfactants. Several mechanisms have been proposed to explain their effect \cite{pugh_foaming_1996, bergeron_polydimethylsiloxane_1997, denkov_mechanisms_1999, garrett_defoaming_2015} and are schematised in Fig.\ref{fig:schema_spreading_bridging}. 
In the case of the spreading mechanism (Fig.\ref{fig:schema_spreading_bridging}a), a droplet enters the air-liquid interface and, because of the surface tension difference between the thin film and the droplet, the droplet spreads provided that the spreading coefficient $S$ is positive. $S$ is defined as :

\begin{equation}
\label{eq:def:S}
S=\gamma_{c}-(\gamma_{d}+\gamma_{d-c})(h_d)
\end{equation} 

with  {$\gamma_c$ and $\gamma_d$ the surface tension at the air interface of the continuous and dispersed phase and $\gamma_{d-c}$, the surface tension between these two phases. In Eq.\ref{eq:def:S}, we suggest that the energy of the spread droplet $\gamma_{d}+\gamma_{d-c}$ could depend on its thickness $h_d$ through long range interactions through its interfaces.} As the droplet spreads, it drags the underlying liquid in the thin film which causes the local thinning of the film. The mechanism is thus driven by the positive spreading coefficient $S$ of the droplet on the thin film. 
The other mechanism described in the literature postulates that a droplet bridges the film, meaning that it enters both air-liquid interfaces. Again, due to the surface tension difference, the droplet, which is a deformable object, stretches and thins down, causing the film's bursting. The bridging coefficient $B$, defined as $B=\gamma^2_{c}+\gamma^2_{d-c}-\gamma^2_{d}$, has been introduced to assess if the phenomenon is thermodynamically favorable\cite{garrett_preliminary_1980}. If $B>0$, the bridge is unstable and vice versa. Moreover, the force per unit length driving thinning is larger than twice the spreading coefficient, which suggests that this mechanism is faster than the spreading mechanism.
We note that both mechanisms require the entry of the droplet at the air-liquid interface. This turns out to be a critical step in aqueous foams for which surfactants can introduce an energy barrier. We have shown\cite{delance_uptake_2022} that in surfactant-free system considered herein, the droplets enter without delay at the air-liquid interface. Moreover, we have measured positive values of both $S$ and $B$ with $S=4.4\pm0.7~$\SI{}{\milli\newton\per\meter} and $B=254\pm 1~$\SI{}{\square\milli\newton\per\square\meter}, meaning that both spreading and bridging mechanisms are favorable. Yet, we will show next that {\it spreading} and {\it bridging} differ by the size of the depressions formed and their characteristic time.

\begin{figure}
	\centering
	\includegraphics[width=0.8\linewidth]{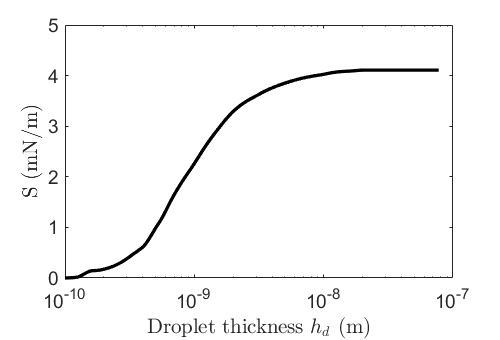}
	\caption{Spreading coefficient $S$ defined by Eq.~\ref{eq:def:S} as a function of the thickness $h_d$ of a microdroplet spreading at the air-oil interface. $S$ decreases as a droplet thins down. Droplet radius $R_d$=\SI{1.2}{~\micro\meter}, viscosity $\eta_d$=6~Pa.s, PDMS chains molecular diameter $d=0.7$~nm and radius of gyration $8$~nm. For $h_d>d$, the spread droplet is dense and $h_d$ stands for its thickness. For $h_d<d$, $h_d$ stands for a surface density. 
	}
	\label{fig:S_h_drop}
\end{figure}

In the following, we suppose that the radius of the depression we observe (Fig.~\ref{fig:goutte_spreading}b,c) is equal to that of the spread or stretched droplet. In addition, we make the hypothesis that once spread or stretched, the droplet, whose initial radius is $R_d$, is a disk of thickness $h_d$ and radius $R$. Thus, using volume conservation, we get:
\begin{equation}
\frac{4}{3}\pi R_d^3=h_d\pi R^2.
\label{eq:volume_conservation}
\end{equation}
In a {\it spreading} mechanism, the spreading of a droplet stops when the spreading coefficient $S$ becomes zero, and the enhanced thinning of the film stops simultaneously. 
At this stage, the center thickness of the depression may or not have reached the critical thickness $h^*_{min}$ for bursting. If not, the depression formed by the spreading of the droplets heals because of capillarity and finally reabsorbs as observed in Fig.\ref{fig:disparition_goutte}. From a previous work on the spreading of microdroplets at the oil-air interface\cite{delance_uptake_2022},  {a measurement of the energy of the spread droplet $(\gamma_{d-c}+\gamma_{d})(h_d)$ as a function of its thickness $h_d$ was obtained. More precisely, $h_d$ stands for a thickness when it is larger than a molecular diameter $d$ so that the dense hypothesis holds. For PDMS, $d$ can be estimated as the diameter of a polymer chain lying flat at the interface \cite{schune:rouse:2020} and $d=0.7$~nm. For comparison, the radius of gyration is 8~nm. Below this molecular size, $h_d$ stands for a surface density or a volume per unit area and the PDMS is no longer dense. Using Eq.~\ref{eq:def:S},} we compute the variations of the spreading parameter $S$ with $h_d$ in Fig.\ref{fig:S_h_drop}. We find that $S$ is divided by a factor of 2 when $h_d$ decreases from $10$~nm down to $\approx 1~$nm, which means that for $h_d \leq 1$~nm, the thinning of the depression is slowed down until it stops: the depression resorbs under the action of capillary forces.
Thus, considering an initial droplet radius $R_d\sim$ \SI{1.2}{\micro\meter}, a final thickness ranging between $h_d\sim1$ and 10~nm, we expect from Eq.\ref{eq:volume_conservation} that the radius of the depression in a spreading mechanism $R_{spreading}\sim \frac{R_d^{3/2}}{h_d^{1/2}}$ ranges between $\sim10$ and 50~$\mu$m.  {This scaling also shows that the extension of the spread droplet $R_{spreading}$ will be all the more larger than the initial microdroplet radius $R_d$ is large.}

In contrast, the {\it bridging} mechanism is irreversible because the droplet/air interface tension $\gamma_d$ remains always smaller than the film tension $\gamma_c$: the droplet keeps stretching until it reaches  {its rupture thickness $h_{min}^\star$. We assume that the order of magnitude estimate of $h_{min}^\star$ barely depends on the film composition, that is, on the thinning mechanism, so that $h_{min}^\star$ is taken around 25 to 50~nm consistently with Figs.~\ref{fig:phase_continue} and \ref{fig:goutte_spreading}c. Using the volume conservation equation (Eq.~\ref{eq:volume_conservation}) with $h_d=h_{min}^\star$ and $R_d=1.2~\mu$m, we find that the radius $R$ of the droplet cannot be larger than $R_{bridging}\leq$\SI{10}{\micro\meter}}.  {Finally, the surface energy gain driving the stretching of bridges is $S_b=2(\gamma_{c}-\gamma_{d})\sim 10$mN.m$^{-1}$, more than twice larger than $S$ so that bridging is expected faster than spreading}. Altogether, we demonstrate that the spreading and bridging mechanisms exhibit different depression's radii as well as different time scales that will help us discriminate between them.

\begin{figure}[h!]
	\centering
	\includegraphics[width=0.8\linewidth]{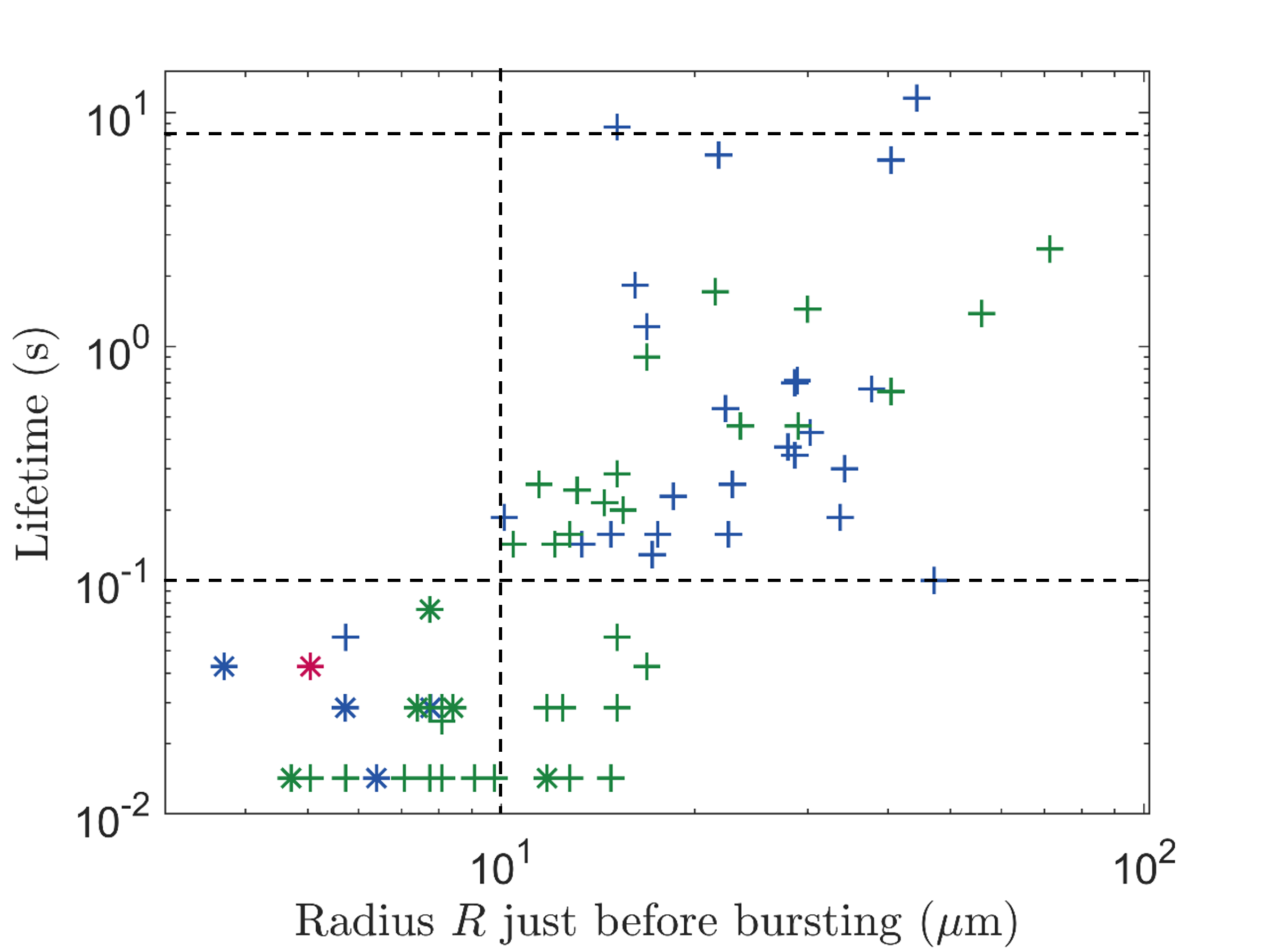}
	\caption{
		Lifetime of suspended films containing microdroplets as a function of the radius $R$ of the thinning zone right before bursting for different droplet viscosities $\eta_d$=27~Pa.s (green), 6~Pa.s (blue), 0.3~Pa.s (red). Droplet initial radius $R_d$=\SI{1.2}{~\micro\meter}. Same markers as Fig.~\ref{fig:distributions}. Radii $R<10~\mu$m correspond to lifetimes smaller than 0.1~s consistently with bridging; Lifetimes larger than 0.1~s correspond to $R>10~\mu$m consistently with spreading.}
	\label{fig:radius_depression}
\end{figure}

\begin{figure*}
	\includegraphics[width=0.9\linewidth]{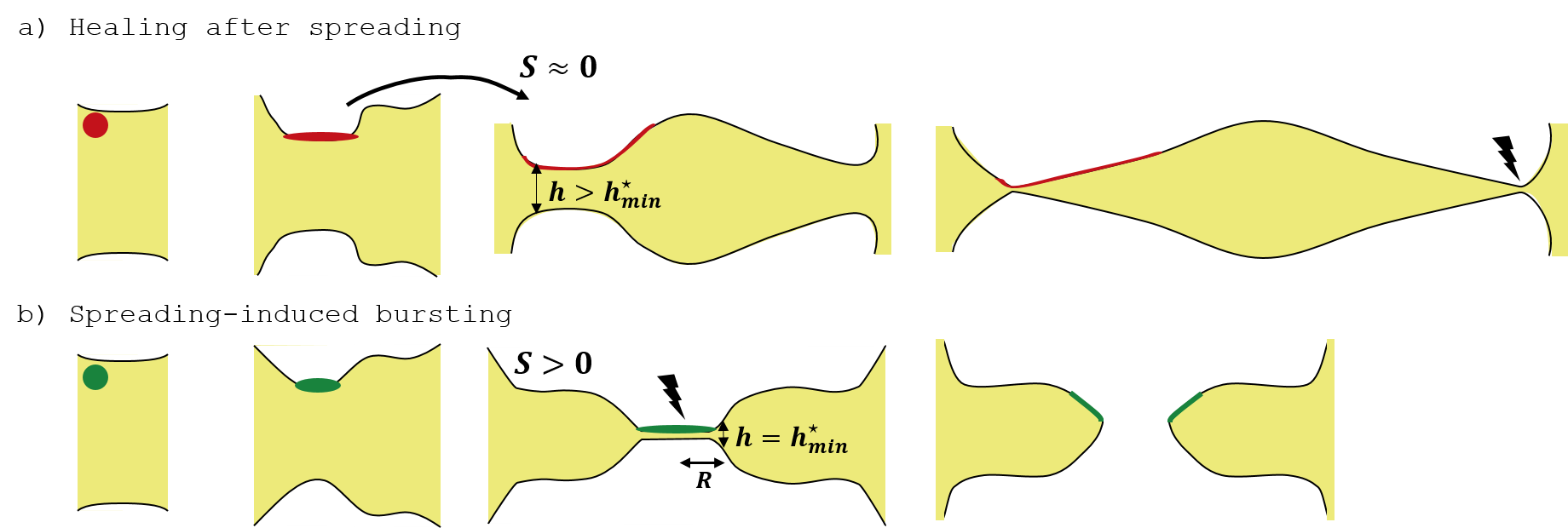}
	\caption{Schematic representation of the effect of droplet viscosity on the thinning of liquid films. (a) For the droplets with the smallest viscosity, spreading time is much lower than the drainage time of the film. As a consequence, the spreading of the droplet is statistically not long enough to reach the critical thickness for bursting and the depression often heals. (b) On the contrary, for droplets with the largest viscosity, the spreading time of the droplet is longer so the local thinning is also slower, but this gives more time for the film to drain until it reaches the critical thickness. The depression radius right before bursting is $R$.}
	\label{fig:schema_viscosite}
\end{figure*}

Now, we measure the radius of the spread droplets just before the film bursting, denoted $R$ and defined in Fig.~\ref{fig:schema_spreading_bridging}. In Fig.~\ref{fig:radius_depression}, we plot the film lifetime as a function of $R$. \\
Firstly, we observe that, for more than 95\% of the depressions, the radius $R$ is larger than \SI{10}{\micro\meter}, which points to a {\it spreading} mechanism occuring in most cases (see Fig.\ref{fig:goutte_spreading}). {The size of the depression may even reach several tens of microns, which is inconsistent with a bridging mechanism and thus further evidences a spreading mechanism. Since it is a reversible mechanism, droplet spreading is also consistent with observations of the healing of depressions (see Fig.\ref{fig:disparition_goutte}). In contrast, 
	we note that the rare cases for which $R<$\SI{10}{\micro\meter} are associated with lifetimes lower than 0.1~s (see an example in Fig.\ref{fig:goutte_bridging}): we suggest these are cases of
	a bridging mechanism, in which 
	the droplet enters the film early, before the formation of a dimple, and causes its quick bursting. At this stage, the diameter of the droplet is indeed similar to the thickness of the film. Bridging actually causes an unbalanced tension in the film that drives its stretching with a fast plug flow. Tensions balance, which is necessary for the development of a slower Poiseuille flow, as described in thin films with surface elasticity \cite{lhuissier_bursting_2012, tregouet_instability_2021, tran_mechanisms_2022}, cannot be established. As a consequence, we expect that bridging cannot give way to the formation of any dimple, which is associated with a velocity gradient  {across} the film thickness. Conversely, we observe the formation of a sub-dimple in films with large lifetimes (Fig.\ref{fig:goutte_spreading}c), which we then link to a Poiseuille flow induced by the drop {\it spreading}.} \\
{Hence, in the following, lifetimes $T$ between 0.1~s and 8~s will be ascribed to a spreading mechanism. From Fig.~\ref{fig:distributions}b)-insert, we observe that spreading is more efficient at reducing the film lifetime when the microdroplet radius $R_d$ is larger. This observation is consistent with Eq.~\ref{eq:volume_conservation}, using $h_d=1$~nm as the limit for which spreading stops, which predicts that $R_{spreading}$ increases with $R_d^{3/2}$.}\\
We eventually show that, in our system, the microdroplets destabilise the films via both spreading and bridging mechanisms. The latter is rare but reduces the film's lifetime efficiently: it is fast  {($T<0.1~s$)} and irreversible. The former occurs much more often but is slower  {($0.1<T<8$s)} and reversible: healing is observed.  {More importantly, {\it spreading} is all the more efficient than the radius $R_d$ of the microdroplets is larger.} Altogether, our results allow us to ascribe the anti-foaming effect to a predominant spreading mechanism. In the following, we investigate the effect of the droplet viscosity and, more generally, we emphasize the key role of the moment of entry and the spreading velocity of the microdroplet on the efficiency of the spreading mechanism to induce film bursting.

\subsection{Entry and spreading times at the air-liquid interface}
In this section, we explain why some droplets fail to cause the film to burst, in particular why some depressions heal and why droplets with the smallest viscosity have no effect on the film lifetimes, as seen in Fig.~\ref{fig:distributions}, while only a few depressions can be seen that do not lead to film bursting.
As stated in the previous section, a depression can heal if the droplet is fully spread before it reaches the critical thickness (see an example in Fig.\ref{fig:disparition_goutte}). This highlights that the droplets induce the bursting of films only when the latter are already thin enough. In other words, the spreading time of the droplet $\tau_{spreading}$ should be of the order of magnitude of the drainage time of the film $\tau_{drainage}$. If  $\tau_{spreading}\ll \tau_{drainage}$ and the droplet enters early at the air-liquid interface, that is, when the film is still several microns thick, the resulting depression will not be deep enough to cause the film's bursting. As schematised in Fig.\ref{fig:schema_viscosite}, the values of entry and spreading times are crucial to reduce the film's lifetime, a droplet should not enter too early at the air-liquid interface. While the entry of the droplet is driven by a diffusion-convection which can hardly be tuned independently from the other parameters, its spreading time and velocity strongly depends on its viscosity. Indeed, in our experiments, the droplets are between 40 and 4000 times more viscous than the continuous phase.  {Brochard-Wyart \textit{et al.}\cite{brochard-wyart_spreading_1996} modelled the case of a viscous droplet spreading on a liquid bath with negligible gravity: the spreading velocity $\dot{R}$ varies as the ratio of the spreading parameter $S$ to the extensional viscosity $\eta_e$ of the drop: $\dot{R}\sim S/\eta_e$.  Thus, while keeping the droplet's initial radius constant, increasing the PDMS molecular mass increases both the shear viscosity and the extensional viscosity $\eta_e$} and causes an increase of the spreading time. Note that other spreading dynamics were found when the viscous shear in the film is the limiting dissipative mechanism \cite{vernay_bursting_2015}.  {Extensional viscosity is known to be difficult to assess but its ratio to the shear viscosity was measured constant in PDMS melts \cite{chatraei_lubricated_1981} so we offer next to compare the PDMS grades between them}. Hence, we understand the effect of PDMS viscosity on the film's lifetime as follows. In the case of the least viscous droplets (red curve), we assume that $\tau_{spreading}\ll \tau_{drainage}$. As a consequence, most of the droplets don't cause the bursting of the film which heals (Fig.~\ref{fig:schema_viscosite}a). On the contrary, in the case of the most viscous droplets (in green), the spreading is slower, which gives more time for the film to drain and reach $h_{min}^\star$ where the film bursts, as illustrated in Fig.~\ref{fig:schema_viscosite}b).\\

\section{Conclusions}

In this paper, we studied the influence of dispersed microdroplets on the stability of suspended films composed of liquid mixtures. We show that a careful choice of system allows us to effectively compare the lifetime distributions of suspended films with and without microdroplets. The monophasic reference state corresponds to long lifetimes and can be rationalised within the framework of the Gibbs' elasticity of their interface with air which was earlier discussed in the literature \cite{tran_mechanisms_2022}. From this, by comparison, the biphasic case can here be discussed at the scale of the film and of the microdroplets of the emulsified phase. Two mechanisms by which microdroplets accelerate the film bursting are clearly identified in both the lifetimes distribution and the time evolution of the thickness profiles of the films. Indeed, we observe that the droplets generate the formation of local depressions in the film, which, in some cases, lead to its bursting. We show that most of these depressions are due to the spreading of a droplet at the surface of the film, which accelerates the drainage of the underlying film. In rare cases, a bridging mechanism occurs and causes the fast bursting of the film. 
{
	Since our analysis relies on volume conservation and extensional flow within the microdroplets but not on the film interfacial properties, providing that the spreading parameter is positive, we believe that our results regarding the effect of microdroplet size and viscosity can be extended to aqueous systems with surfactants or any other types of systems: microdroplets are all the more efficient that they are more viscous and larger. Indeed, both viscosity and microdroplet size tend to lengthen the spreading of microdroplets at the air interface and to give more chance to the film to thin down to the critical thickness required for bursting.
	Also, thinning by bridging will always be faster than thinning by spreading, whatever the system under consideration.
	In contrast, the relative occurrence of bridging versus spreading should be specific to the system.  It depends on the Entry parameter and on the ratio between the droplet size and the film thickness, the latter being strongly dependent on the viscosity and interfacial behaviour of the continuous phase as suggested \cite{denkov_mechanisms_1999} in the past.}

Our results offer a basis to quantitatively assess the anti-foaming effect of these microdroplets in liquid mixtures as a function of their size and viscosity.


\section*{Conflicts of interest}
There are no conflicts to declare.

%

%
%

%
\bibliographystyle{rsc}
\bibliography{bibli_bike_wheel}

	\end{document}